\documentclass[aps,pra,twocolumn,showpacs]{revtex4}%
\def\BibTeX{{\rm B\kern-.05em{\sc i\kern-.025em b}\kern-.08emT\kern-.1667em\lower.7ex\hbox{E}\kern-.125emX}}
\usepackage{graphicx}
\graphicspath{ {./images/} }
\usepackage{color}
\usepackage{bm}
\usepackage{comment}
\usepackage{amsmath, amsthm, amssymb,mathrsfs}
\usepackage{csquotes}
\usepackage[colorlinks=true, citecolor=blue,allcolors=blue]{hyperref}
\newcommand{\Schro}{Schr\"{o}dinger}

\begin{document}
\title{A Science Gateway for Atomic and Molecular Physics}

\author{Barry~I.~Schneider}
\email{barry.schneider@nist.gov} 
\affiliation{National Institute of Standards and Technology, Gaithersburg, Maryland 20899, USA}
\author{Klaus~Bartschat}
\email{klaus.bartschat@drake.edu}
\author{Oleg Zatsarinny}
\email{oleg_zoi@yahoo.com}
\affiliation{Drake University, Des Moines, IA 50311, USA}
\author{Igor Bray}
\email{igor.bray@curtin.edu.au}
\affiliation{Curtin University, Perth, GPO Box U1987, Western Australia}
\author{Armin Scrinzi}
\email{armin.scrinzi@lmu.de}\affiliation{Ludwig-Maximilians Universit\"at, M\"unchen, Germany}
\author{Fernando Mart\'in}
\email{fernando.martin@uam.es}
\author{Markus Klinker}
\email{markusklinker@gmail.com}
\affiliation{Universidad Aut\'onoma de Madrid, Catoblanco, Madrid 28049 Spain}
\author{Jonathan Tennyson}
\email{j.tennyson@ucl.ac.uk}\affiliation{Department of Physics and Astronomy, University College London, Gower Street, London, WC1E 6BT United Kingdom}
\author{Jimena D. Gorfinkiel}
\email{jimena.gorfinkiel@open.ac.uk}\affiliation{School of Physical Sciences, The Open University, Milton Keynes MK7 6AA, United Kingdom}
\author{Sudhakar Pamidighantam}
\email{pamidigs@iu.edu}
\affiliation{Indiana University, CIB 2709 E 10th Street, Bloomington, In 47408 and the eXtreme Science and Engineering Discovery Environment(XSEDE)}
\date{\today}


\begin{abstract}
We describe the creation of a new Atomic and Molecular Physics science gateway (AMPGateway). \\

Note:This paper was supposed to appear in an ACM transactions of PEARC19.  It was accidentally omitted. \\

The gateway is designed to bring together a subset of the AMP community to work collectively 
to make their codes available and easier to use by the partners as well as others. 
By necessity, a project such as this requires the developers to work on issues of portability, 
documentation, ease of input,  as well as making sure the codes can run on a variety of architectures. 
Here we outline our efforts to build this AMP gateway and future directions. 
\end{abstract}
\maketitle

\section{Introduction}
On May 14-16, 2018, an NSF supported workshop entitled, 
``Developing Flexible and Robust Software in Computational Atomic and Molecular (A\&M) Physics'' 
was organized by Barry Schneider (chair), Robert \hbox{Forrey} (Penn State) 
and Naduvalath Balakrishnan (UNLV) at the Institute for Theoretical Atomic and Molecular Physics, 
Harvard-Smithsonian~\href{https://www.cfa.harvard.edu/itamp-event/developing-flexible-and-robust-software-computational-atomic-and-molecular-physics-0} {ITAMP}~\cite{ITAMP}.  
The purpose of the workshop was to bring together a group of internationally known researchers in computational atomic and molecular physics to:
\begin{itemize}
\item Identify and prioritize  outstanding problems in A\&M science, which would benefit 
from a concerted community effort in developing new software tools and algorithms that would 
lead to more rapid and productive scientific progress for the entire community.
\item Discuss approaches to optimize achieving that goal.
\item Produce and disseminate a report of the workshop to the community.
\end{itemize}
\newpage
A concerted community effort is underway to develop and maintain these tools in order to ensure 
continued scientific progress. The group acknowledged that, in contrast to some other communities, 
A\&M physics has lagged behind in developing community software that is robust and can be used, in a 
relatively straight\-forward way, by other 
than the group who developed that software.  While there are exceptions, many software packages are poorly documented, 
poorly written, and only usable by a set of local ``experts''.  The tools themselves are capable of 
treating scientific and technologically interesting problems, but they are only accessible to a small 
group of people.  The codes are not always maintained, and the lack of coordination among the 
developers has led to a lot of ``reinventing the wheel''. The group felt strongly that the efforts 
being expended in developing these computational tools should be available and usable by 
future generations of A\&M scientists.

The success of the workshop led six of the groups to work together and develop an XSEDE proposal 
to build and maintain a Science Gateway devoted to the codes developed in these groups.  
That proposal was supported, and since May of 2018 there has been decent progress.  A number of the 
codes are already ported and running on various XSEDE platforms. Some progress has 
been made in making them usable by others within the group but not yet the outside world.  
We are now taking steps to achieve this last goal. 

The AMPGateway uses the multi-tenanted Apache Airavata middleware 
framework~\cite{scigap,airavata,keycloak} served by the SciGaP hosting services for sustained operation.
In the first stage of our efforts the software suites have been deployed as 
independent applications with specific input interfaces. Community building 
has already started and a few additional software suites have been identified 
for inclusion in phase two. The interoperability of the  software suites is 
very important and  will be addressed as a follow-on.

The present manuscript is 
divided into four major sections. In the Introduction we provide a history of how and 
why the project got started and our decision to go to XSEDE~\cite{xsede}  for support 
for the gateway. In Section~II we present some information on the AMP codes that are 
already available on the gateway. 
Section~III is devoted to the details of the construction and deployment  of the gateway. 
In Section~IV we discuss issues of broadening usage of the gateway and questions of community building.  

\section{Current Code Status}
At present, we have concentrated our major effort on five codes. A brief description of these
packages is given below.
\subsection{BSR}
The $B$-spline $R$-matrix (BSR) method and the accompanying computer code~\cite{BSR3} were developed 
by Oleg \hbox{Zatsarinny} in the group of Klaus \hbox{Bartschat} at Drake \hbox{University}.  The program  
computes transition-matrix elements for electron collisions with atoms and ions as well as photo\-ionization
processes. From these, cross sections and other experimentally observable parameters can be obtained.
The code can also be run in a mode that provides atomic structure information
through energy levels and oscillator strengths.

The BSR approach is a particular variant of the $R$-matrix method to solve the 
close-coupling equations in coordinate space. In this respect, it is complementary to the convergent close-coupling 
(CCC) approach described below.  BSR is an alternative formulation of the well-known
$R$-matrix code developed in Belfast under the long-term leadership of Philip~Burke.
The Belfast code is somewhat singular in that it is readily available and used by a small group of users.  While the last general write-up appeared in 1995~\cite{BEN1985}, 
updated versions are available~\cite{Badnell}.  A comprehensive introduction to $R$-matrix theory 
for atomic and molecular collisions processes, as well an overview of many applications,
can be found in the book by Burke~\cite{Burke2011}. 

The published BSR code~\cite{BSR3} is a serial version, which was written in the non-relativistic 
and semi-relativistic (Breit-Pauli) frameworks.  Relativistic (DBSR) and MPI-parallelized versions,
as well as extensions to treat ionization processes (similar to the CCC method described below)
exist and are being used by the developer and a small group of collaborators.  Executables of the parallelized codes
(currently running on Stampede2) will be uploaded to the Gateway in the near future.
The BSR and DBSR packages are a prime example 
where updated documentation
and a wide distribution are urgently needed before critical expertise is lost. Fortunately,
the urgency was recently recognized by the NSF and resulted in the funding of a three-year proposal to achieve
exactly these goals.  We expect the gateway described in the present paper to be one of the vehicles to ensure  
significant future progress.  

A comprehensive overview of the BSR method and its applications at the time was published by Zatsarinny and Bartschat~\cite{ZB2013}.
The most noteworthy features of the code are:
\begin{itemize}
\item Use a finite-element ($B$-spline) rather than a finite-difference approach  in the 
      calculation of the matrix elements needed to set up the hamiltonian in the inner region.
\item Employ non-orthogonal sets of one-electron orbitals to account for the term-dependence of 
      the valence orbitals, in particular for complex, open-shell targets, thereby 
      providing an economical and accurate description of the target states and much 
      flexibility in building the scattering wavefunction as well as pseudo\-states to 
      further improve the target description and enable the treatment of electron-impact single-ionization 
      as well as photon-driven double ionization processes.
\end{itemize}
The BSR code has the following major parts: 
\begin{itemize}
\item Build the $N-$ and $(N+1)-$electron configurations.
\item Generate all necessary one-electron and two-electron matrix elements to set up the
      target and scattering Hamiltonians in the internal region.
\item Diagonalize these Hamiltonians.
\item Propagate the wavefunction from the $R$-matrix boundary, $r=a$, to ``asymptotia'' ($r_b$), where it can be matched 
to known analytic forms. The propagation requires the solution of a set of coupled differential equations using 
known long-range potentials and needs to be repeated for each scattering energy. If angle-differential ionization processes with two free electrons
      in the final state are to be treated as well, the inner region may need to be increased beyond the
      original criterion.
\end{itemize}
Even though there is no general way to predict where most of the computational effort is needed,
in most cases the generalized eigenvalue problem (diagonalization with {\it all} eigenvalues needed) 
of the $(N+1)-$electron Hamiltonian is a very time-consuming step. For complex targets, setting up this 
Hamiltonian can be expensive as well.  For ionic targets, the wealth of resonances may 
require many thousands of collision energies to be treated, which can result in significant time
going into the asymptotic region.

To summarize:  The BSR method is closely related to both the Belfast $R$-matrix approach and the 
CCC method described below. The two $R$-matrix codes were designed to handle complex
targets and many energies, while the CCC code can handle more processes but is essentially limited to
quasi-one and quasi-two electron targets. Some benchmark comparisons for problems that all three methods
should be able to handle have been performed.  As expected, the results are numerically equivalent, 
but one or the other method may be more efficient.  The details strongly depend on the complexity of the target and the
energies for which results are required.

\subsection{CCC}
The original implementation of the Convergent Close-Coupling method was designed
to produce accurate cross sections for scattering of light projectiles from quasi
one- and two-electron targets~\cite{BS95b}. It began with
electron-hydrogen scattering~\cite{BS92}, and was further extended to
quasi-one electron targets well-modeled by one valence electron above
a frozen Hartree-Fock core~\cite{B94}. It was then extended to the
helium target~\cite{FB95}, and quasi two-electron targets such as Be~\cite{FB97}.
The main features are:
\begin{itemize}
\item Expansion of the target state in a sufficiently complete
  $\mathscr{L}^2$ basis size $N$ to treat cases where both excitation and
  ionization of the target are possible, with convergence tested by
  systematically increasing $N$.
\item Expansion of the scattering wavefunction in the momentum based Lippmann-Schwinger equation.
\item Introduction of numerical quadrature to reduce the problem to a very large set of linear algebraic equations.
\end{itemize}

CCC has been extended to positron scattering, where the positron introduces 
a second center capable of forming the electron~positron bound state known as positronium (Ps)~\cite{KB02}. 
This is an example of a rearrangement collision and as such introduces 
even more complexity into the computational procedure.  Such calculations can be ``time-reversed'' 
to be considered as Ps scattering on (anti)protons to form
(anti)hydrogen~\cite{Kadyrov15}. A review of the CCC method for
positron scattering has been given by \citet{KB16}.

On the computational side, the CCC codes have been parallelized to use OpenMP on a node and MPI 
between nodes and have been deployed successfully on Comet and Stampede2. A GPU implementation, currently underway, shows
immense promise with up to two orders of magnitude speedup.

\subsection{UKRMol(+)}

The UK Molecular R-matrix codes were  designed to treat low-energy elastic and inelastic electron-molecule collisions
using the R-matrix method; they have evolved  to also study photoionization and positron-molecule collisions as well as to 
produce the input required for time-dependent molecular R-matrix with time dependence (RMT) calculations \cite{RMT}. Similar to BSR they 
are based on the R-matrix method and the general theory can be found in the book by Burke \cite{Burke2011}. 

The now frozen release version of the (mainly serial) UKRMol suite uses Gaussian Type Orbitals (GTOs) to represent
both the target and continuum orbitals. A publication presenting this code by Carr {\it et al.}
\cite{jt518} followed a project which substantially updated (to Fortran95) and standardized the programming used, particularly 
in the inner region. A review article by Tennyson \cite{jt474} from the same period gives a comprehensive overview of 
theory used and the functionality of the code. 

The use of GTOs to represent the continuum leads to constraints on both the size of the inner region
that can be used and the free electron energy range. Recently a new suite, known as UKRMol+, has been developed led 
by Zden\v{e}k Ma\v{s}\'{i}n and Jimena Gorfinkiel~\cite{jt518}. The code uses the new GBTOlib integral library to 
determine all the one- and two-electron integrals needed in the mixed basis of GTOs  and B-splines: it offers the choice of using GTOs, B-splines or hybrid
GTOs -- B-splines to represent the continuum; the bound orbitals are always described by GTOs. The library, 
which uses object oriented features from the Fortran2003 standard, involves distributed and shared-memory parallelization.
UKRMol+ incorporates a number of algorithmic improvements, including a faster configuration state function (CSF) 
generation and parallization of the construction and diagonalization of the $N$ and $(N+1)$ Hamiltonians \cite{ukrmol+}. 
Further parallelization to avoid I/O to disk during the evaulation of transition moments for photoionization and RMT input is currently being tested. 

Both suites contain a rich array of outer region
functionality including automated resonance detection and fitting, bound state detection, computation of multichannel 
quantum defects, rotational excitation and evaluation of photoionization cross sections. So far applications of the UKRMol+ suite
are limited \cite{jt682,rmatrix_photo,rmatrix_alanine}, but a full release and associated article will be available shortly \cite{ukrmol+}.

The codes have been available as freeware for more than a decade and are widely used: the software can be downloaded 
as a tarball and installed (in the case of UKRmol+) using  cmake, provided the necessary libraries are available in the system. 
Neither suite is straightforward to use without training; Quantemol-N \cite{jt416} is a commercial front end which has 
led to a significant increase in the user base of the code. A set of perl scripts developed by Karel Houfek that 
simplify writing the input is also now available, both for electron scattering and photoionization calculations. 
Further details can be found on the website of the UK Atomic, Molecular and Optical physics R-matrix consortium (https://www.ukamor.com/). 
 
\subsection{tRecX}
The tRecX code package~\cite{tRecXweb,tRecXgit} is a general framework for solving initial value problems of the form
 \begin{equation}
 \frac{\partial} {\partial t} \Psi = \mathcal{D}[\Psi,t] + \Phi(t)
\end{equation}
for an arbitrary number of spatial dimensions and a variety of coordinate systems. 
The main design is for linear $\mathcal{D}$, but non-linear operators can also be used.  

\subsubsection{Applications}

\begin{figure}[t]
\begin{minipage}[c]{3.5cm}
\includegraphics[scale=0.24]{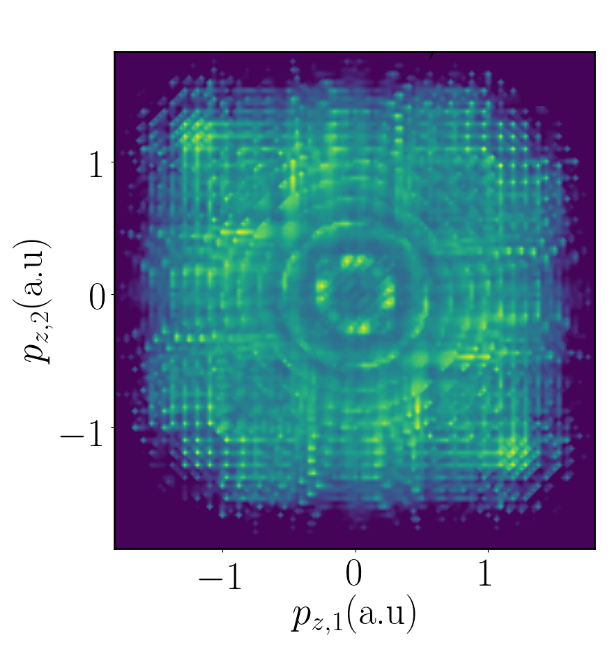}
\end{minipage}
\hspace*{0.3cm}
\begin{minipage}[c]{3.5cm}
\includegraphics[scale=0.12]{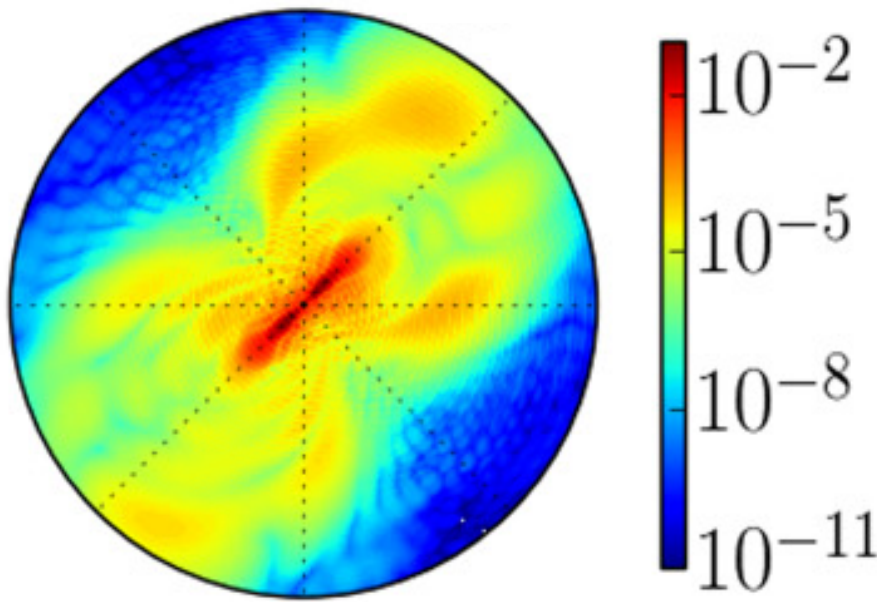}
\end{minipage}
\caption{\label{fig:tRecXexamples}
Left: Helium in full dimensions, double emission cross-section $\sigma(p_{z,1},p_{z,2})$ 
for a 2-cycle pulse at wavelength 400 nm and 5$\rm \times10^{14}\,W/cm^2$ intensity. Anti-correlated emission is favored.
Right: haCC calculation for $\rm CO_2$, photo-emission by an 800 nm laser pulse at intensity $10^{14}\,\rm W/cm^2$ 
up to energies 2.5~au in the $xz$-plane at 45$^\circ$ alignment of the 
molecular axis to polarization direction (from \cite{majety17:co2spectra}.) }
\end{figure}

The code has been primarily used for solving the time-dependent {\Schro} equation of atomic
and molecular systems in ultra-short pulses and in strong near-IR fields. The most significant results are
fully differential spectra for single- and double photo-emission from the He atom at near 
infrared wave-length~\cite{zielinski16:doubleionization}, strong field ionization rates  of noble gases~\cite{majety15:static} 
and differential spectra for small di- and tri-atomic molecules~\cite{majety15:dynamicexchange,majety17:co2spectra}, 
cf.\ Fig.~\ref{fig:tRecXexamples}.

\subsubsection{Methods}
The code uses three newly developed key techniques:
\begin{enumerate}
\item irECS --- {\em ``Infinite range exterior complex scaling''}~\cite{scrinzi10:irecs} as absorbing boundary conditions and for the 
computation of life-times. irECS is a variant of exterior complex scaling with an infinitely wide boundary 
for absorption at all wave lengths.
\item tSurff --- the {\em time-dependent surface flux} method~\cite{scrinzi12:tsurff} for photo-emission spectra. 
By tSurff, the actual numerical solution remains contained in a small reactive region of typically 20 to 100 atomic units.
\item
haCC --- the {\em ``hybrid anti-symmetrized Coupled Channels''} method~\cite{majety15:hacc} combines Gaussian-based neutral 
and ionic CI states with a numerical single-electron basis. The numerical basis extends over the whole 
system, thus ensuring the proper description of energetic electron-molecule collisions. 
\end{enumerate} 
Two of these techniques are reflected in the code's name

\centerline{ tRecX = tSurff + irECS}

\subsubsection{Structure and inputs}
An effort was made to keep the code flexible without sacrificing efficiency.
Systems with dimensions from one (popular models) to six (He in elliptically polarized
fields), as well as multi-channel models (photo-electron spectra for molecules) are treated 
within the same framework:
the degrees of freedom map into a tree hierarchy, inducing
a tree-structure for vectors and operators, and producing transparent and efficient code by recursion.

Basis functions are arranged in a tensor-tree with  
a variety of basis sets, finite-elements, FE-DVR, and grids that can be combined
on any number of coordinate axes, with multiple basis sets on the same axis. 
Discretization and operators can all be input-controlled. For example,
{
\begin{verbatim}
  #define BOX 20
  Axis: name,nCoeff,lower,upper,funcs,order
  Phi,5,,,expIm 
  Eta,3,-1,1, assocLegendre{Phi}
  Rn,60, 0,BOX,polynomial,15
  Rn,20, BOX,Infty,polExp[0.5]
\end{verbatim}
}
\noindent
with \verb"Eta" for $\cos\theta$ defines polar coordinates.
The bases $\exp(im\phi)$ and $P^{|m|}_l(\cos\theta)$ combine to the spherical harmonics up to $l=2$ and
the $r$-coordinate uses 60/15=4 FE-DVR elements on $[0,20]$ with 15 polynomials on each 
and 20 exponentially damped polynomials at the end. An example for operator specification is
{
\begin{verbatim}
  Operator: hamiltonian
  0.5(<d_1_d><1>+<1><d_1_d>+<Q*Q><1>+<1><Q*Q>)
\end{verbatim}
}
\noindent
for  $(\overleftarrow\partial_x\overrightarrow\partial_x+\overleftarrow\partial_y\overrightarrow\partial_y+x^2+y^2)/2$. 
Many standard operators are pre-defined for various coordinate systems.
\subsubsection{Software}
The code is open source hosted on a Gitlab repository~\cite{tRecXgit}. It is written mostly in C++ and linked 
with some Fortran-based libraries. Optionally, functionality can be extended by linking FFTW and GSL.
Standard compilers are gcc and Intel, ports to Windows and Mac's Clang have been successful but are not actively supported.
Compilation is through CMake, and Doxygen documentation is available. Tutorials and further materials are available in the git  and 
from the tRecX website~\cite{tRecXweb}.
 
\subsection{XChem}
XChem \cite{GABS,xchem} is a solution for an all-electron ab initio calculation of the electronic continuum of molecular systems. 
XChem combines the tools of quantum chemistry (as implemented, e.g., in MOLCAS \cite{molcas}) and scattering theory to accurately 
account for electron correlation in the single-ionization continuum of atoms \cite{xchem,xchemne}, small and medium-size 
molecules \cite{xchemn2}. At its core lies a close-coupling expansion combined with the use of a hybrid Gaussian and B-Spline basis set \cite{GABS}.

This approach yields the scattering states of the molecular system via the eigenstates of the close-coupling matrix (CCM). 
While useful in their own right, the full potential lies in using the close-coupling matrix as a starting point for time-dependent calculations. 
In doing so, one may explicitly model the interaction of molecules with ultrashort (attosecond) pulses. The large band widths of such 
pulses lead to the coherent excitation of multiple ionization channels, whose coupling (accurately described in XChem) gives rise to complex phenomena. 

\begin{figure}[t]
\centering
\includegraphics[scale=.7]{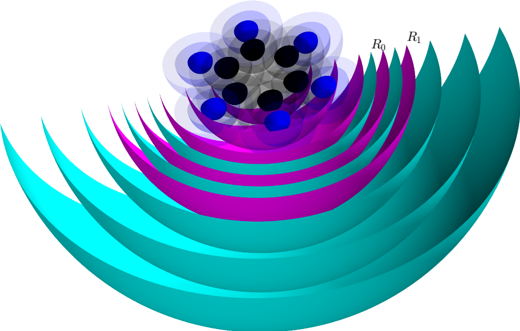}
\caption{Ilustration of the XChem basis architecture in \hbox{benzene}. Cyan: B-splines, mangenta: Gaussians at the center of mass of the molecule, 
blue and black: Gaussians at the atomic sites not overlapping with B-splines.}
\label{fig:GABS}
\end{figure}

An attractive feature of XChem is that the architecture of the basis functions (Fig.~\ref{fig:GABS}) and the use of MOLCAS allow one 
to describe the electronic continuum of medium-size molecules at the same level of theory as multi-reference CI methods do for the 
ground and the lowest excited states of such molecules.  At present the largest systems treated have of the order of ten atoms.

\subsubsection{What can XChem do?}
XChem can compute:
\begin{itemize}
\item The CCM for a user-defined set of ionization channels (each defined as an ionized molecular state coupled to electrons of given angular momenta) 
and including short range states relevant to the problem at hand.
\item Scattering states and scattering phases by asymptotic fitting to the analytical solution.
\item Total and partial photoionization cross sections within perturbation theory. 
\item Lifetime and character of resonances embedded in the molecular continuum, either via analysis of the cross section or 
via inclusion of a complex absorbing potential in the CCM yielding complex eigenenergies.
\item The electron dynamics during and after ionization, caused by and probed with ultrashort laser pulses, by solving the 
time-dependent Schr\"odinger equation using the CCM.
\item The angular distribution of photo electrons (in progress).
\end{itemize}

\subsubsection{What can XChem be used for?}

XChem is a valuable tool for:
\begin{itemize}
\item The theoretical study of ultrafast processes in in Attosecond pump-probe experiments, Attosecond Transient Absorption Spectroscopy (ATAS) 
and Reconstruction of Attosecond Beatings by Interference of Two-Photon Transitions (RABBIT).
\item  The investigation of photoionization of complex molecules close to threshold, where electron correlation effects are crucial to describe the photoionization cross sections.
\item The study of ionization processes intrinsically dependent on electron correlation, like autoionization and Auger decay. 
\item The computation of potential energy surfaces for the investigation of molecular dynamics during and after fast photoionization events.
\end{itemize}

\subsubsection{Who is using XChem?}
\begin{itemize}

\item Researchers in (computational) quantum chemistry or molecular physics interested in studying electron dynamics in the ionization 
continuum of molecules (e.g., photoionization, charge migration, etc).
\item Laboratories investigating ultrafast phenomena in many-electron atoms, small and medium-size molecular systems. 

\end{itemize}

\subsection{Other Interesting Software}
We are also standing up the rather old Many Electron Structure Applications (MESA) 
code that was developed at Los Alamos in the 1980's and modified to compute electron 
scattering and photoionization cross sections from polyatomic molecules using the 
Complex Kohn Method~\cite{mesa}.  While this code is old, and in need of significant modernization, 
it was built to compute electron scattering and photoionization transition matrix elements 
for general polyatomic molecules, a capability not present in other codes and of interest to many users.

There is also an effort underway to incorporate MOLSCAT~\cite{molscat}, a heavy particle 
collision code for vibrational-rotational scattering in molecules. Table~\ref{table:1} summarizes some of the applications deployed.

%
\begin{table*}[t]
\label{table:1}
\begin{center}
 \begin{tabular}{|l|l|l|l|l|l|l|}
 \hline
 \bf Code & \bf Application              & \bf Method             & \bf Parallel & \bf Access                                                             &\bf Restrictions   \\
                   &                                           &                   &    \bf or Serial            &                                 &     \\ 
 \hline \hline
 \bf BSR  & Electron scattering     & R-matrix/B-spline    & S, MPI                          & \href{https://github.com/zatsaroi/BSR3}{BSR (serial)}     & atoms, atomic ions   \\ 
               & Photoionization & & & & \\                                
               & Structure & & & & \\                                
 \hline
 \bf CCC & Electron scattering            & Close-coupling,    & OpenMP  &     & Quasi one- and two- \\
         & Positron scattering                 & Fredholm equations & and MPI &     & electron atoms       \\
          & Photoionization                                          &  in momentum space                                                         &                              &    &  and ions  \\  
 \hline
 \bf UKRMol(+) &Electron scattering,   & R-matrix,  Close-coupling                       & OpenMP &  Public   &  Molecules and    \\
               & Photoionization       & Gaussian and B-spline basis           & and MPI & (\href{https://zenodo.org/}{Zenodo})     & clusters ($\leq$ 30 atoms)                                   \\  
 \hline
\bf tRecX       & Strong-field photo-       & TDSE (tSurff, haCC): grids,  & MPI    & Public                                                        & Small molecules,    \\
                & emission spectra \& rates   & CI-states, FE-DVR, bases     &        &  (\href{https://gitlab.physik.uni-muenchen.de/AG-Scrinzi}{Git}) & two-electron atoms  \\
\hline                    
 \bf XChem       & Scattering states                                   & Close-coupling                                                       &                               OpenMP & Upon request &  Small and     	                         \\
                 & Photoionization                                  & Configuration Interaction   & & & medium-size \\
                 &                       & Hybrid Gaussian and B-spline basis & & & molecules \\
 \hline
\bf MESA        & Electronic structure         &  SCF, MCSCF,                                                 &  S                       & By request       &    Small to                            \\ 
                & Electron scattering           &  CI, Complex Kohn                         &     &    &   medium-size  \\
                & Photoionization   &                            &     &    &    molecules  \\
                                           
\hline 
\end{tabular}
\caption{Some characteristics of the software suites deployed in the AMP Gateway.
}
\end{center}
\end{table*}

\section{AMP Science Gateway Deployment and Application Integration}
The AMP Gateway is deployed using the Apache Airavata Framework.~\cite{airavata}
It relies on the  Science Gateway Platform as a Service (SciGaP) (https://scigap.org/) 
hosting services~\cite{scigap} at Indiana University. The SciGaP platform provides gateway 
services using the  multi-tenanted Apache Airavata middleware. The Airavata core enables 
features such as managing user identity, accounts, authorization provisioning, 
and the ability to access XSEDE and other high performance computational resources 
such as XSEDE's Stampede2, Comet and Bridges.  These resources are transparently 
integrated into the AMP gateway and the batch queues are used for scheduling the 
execution of models using applications and user defined parameters.  

 \subsection{User Accounts, Authentication and Authorization}
The AMP gateway user accounts can be created by users by providing a userid and 
setting a password along with providing email for verification process. 
In addition to this an automated process using the user institutional login 
via CI-Logon~\cite{CILogon} is also provided which avoids the email verification step.  
The gateway administrator controls the access to the resources and needs 
to approve the user for gateway resource access. The users get a ``gateway pending role'' 
when they register. The gateway middleware provides authentication and authorization 
services through the Keycloak~\cite{keycloak} identity management system supported by SciGaP. 
Once the gateway administrator provides approval, the role of the new user will be 
changed to ``gateway-user'' which enables access to gateway resources and applications. 
The gateway additionally provides ``admin-read-only'' and ``admin roles'' with their 
associated permissions for a group to share the admin responsibilities and reporting purposes. 
A  ``gateway-user'' then, can use gateway services such as creating, monitoring, sharing 
and cloning experiments (computational simulations). The users can also add their 
own compute resource allocations using the  functions available under ``User Settings''. 
The ``admin''  users have the authorization to control metadata for accessing XSEDE 
through the gateway ``community login'',  register and deploy applications and their (user) interfaces, manage users, 
and monitor and access all user experiments. These privileges enable 
the admin to efficiently troubleshoot any issues relating to the user services. 
The ``admin-read-only'' users can view all ``admin'' related information but not modify any of the settings.  

\subsection{AMP Application Deployment and User Interface Creation}
\begin{figure}[h]
\centering
\includegraphics[scale=.4]{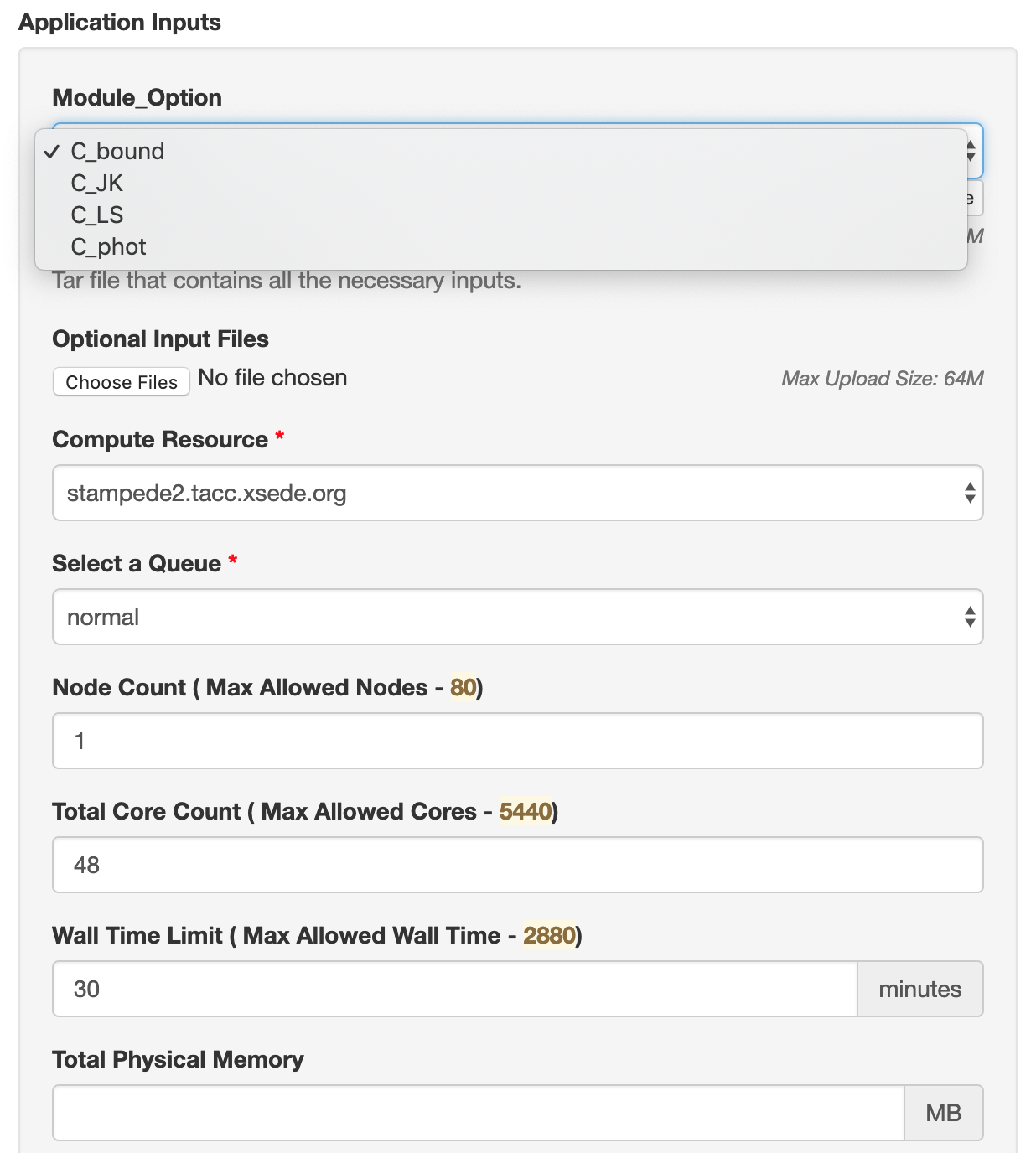}
\caption{\label{fig:fig3} Input interface for BSR software to select a specific module 
for execution and provide inputs as a tarball and set the resource requirements}
\end{figure}

The AMP gateway started with four specific applications described 
above: BSR, XChem, CCC and tRecX suites. These applications were compiled and 
tested on a number of the XSEDE resources. Each of them requires a different 
set of libraries and in the case of XChem integration with other Open source software such as OpenMolcas~\cite{openmolcas}. 
They were independently tested by the collaborating XSEDE ECSS consultant when deployed by the developers or 
deployed by the consultant to establish the 
required environments and tweak make/cmake data. The Application deployment 
in the AMP Gateway consists of defining the application as a ``module'', 
an ``interface'' is defined to user interaction with the application and 
a resource specific ``deployment'' description to fully define it on the gateway. 
The user interfaces are tailored to each application and enable the users 
to provide input parameters using files or variables that either can be sent as 
arguments to the application or a wrapping script. Currently, the interface generator, 
the PHP Airavata client API, provides ways to define file URIs, 
variables (strings, real/integer/Booleans) for inputs and 
standard Error/Outputs and file and variables for outputs. 
Multiple choices for different application sub-modules can be provided to specify 
 and invoke a specific component of the application(for example, Bound/Photonics/LS/JK under software BSR). 
This  can be defined using a simple comma delimited set and can be used in a wrapper 
to pass it as arguments or specified as input parameters for the application as depicted in Fig.~\ref{fig:fig3}. 
The interfaces deployed for the four applications will be further enhanced with 
additional details for input variables and ingesting file sets as archives and 
whole folders in due course. The job submission interface then enables users 
to define HPC resource details such as the system/machine ID that is automatically 
sorted from the list obtained from the deployment description,  queue/partition and 
allocation registered by the administrator or a user, and job specific details such 
as the number of nodes and processors, time and memory specifications. 
In additions these experiment specification the Apache Airavata framework provides 
a sharing mechanism for jobs/experiments and projects (which are collections of experiments) 
to be shared with collaborators and can be set during the experiment creation (or at any time after). 
This allows collaborative job submission, monitoring and analysis of the results. 

\subsection{Monitoring Job Progress}
\begin{figure}[t]
\centering
\includegraphics[scale=.5]{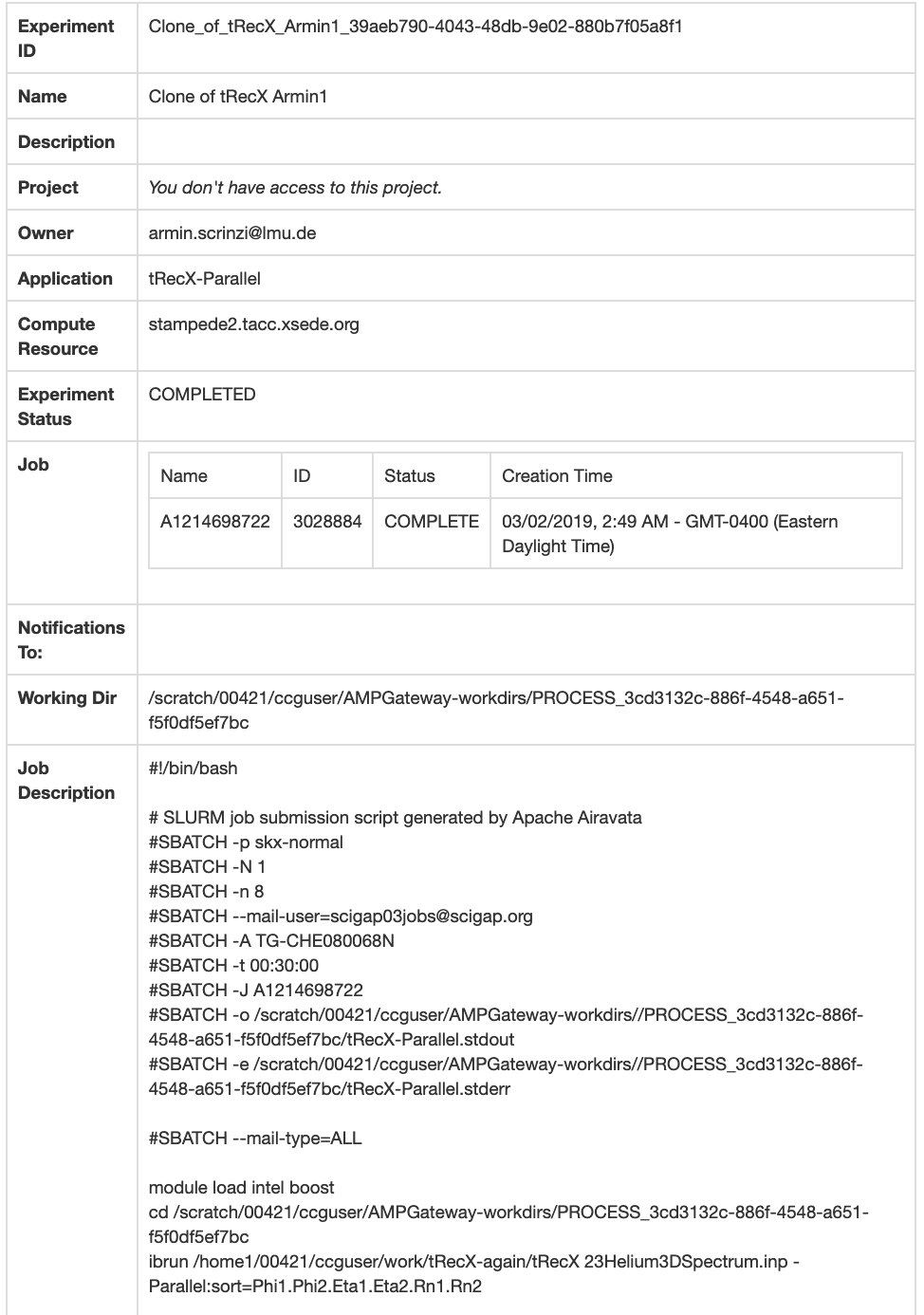}
\caption{\label{fig:fig4} Experiment summary for an job with the batch script created by Airavata middleware}
\end{figure}

\begin{figure}[t]
\centering
\includegraphics[scale=.5]{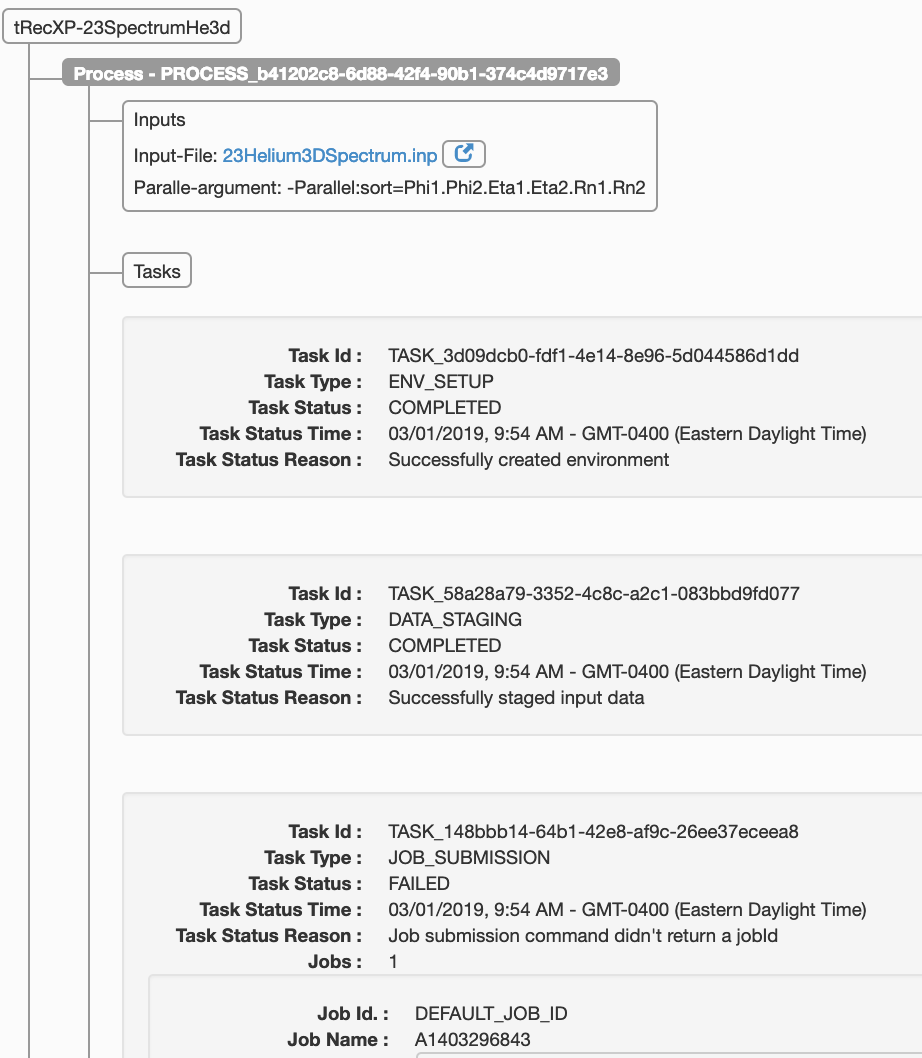}
\includegraphics[scale=.5]{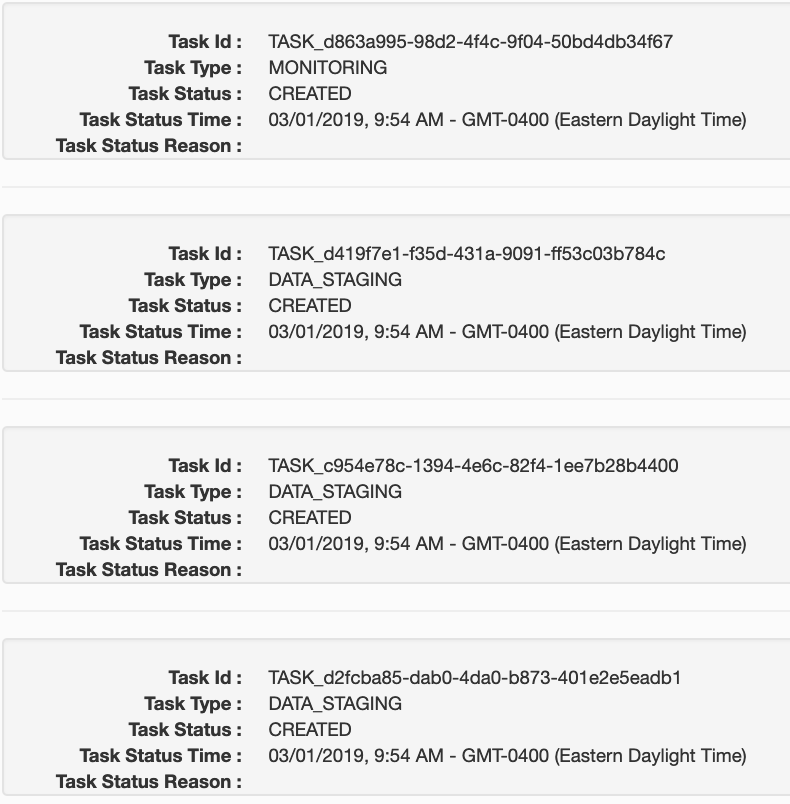}
\caption{\label{fig:fig5} Admin  interface showing task level log for an experiment}
\end{figure}

During experiment creation users can provide their email to receive messages at 
job start and end supplied by the scheduler. Additionally, once the experiment is accepted and launched, 
an ``Experiment Summary'' interface is launched and automatically refreshed periodically 
in order to show the status of the job submitted into the XSEDE resources.  
Currently the status of the job in the scheduler is reflected in the summary interface shown in Fig.~\ref{fig:fig4}. 
Gateway users can monitor experiments owned by them or shared with them by other gateway users. 
Gateway administrators can monitor all gateway experiments using 
the ``Experiment Statistics'' page available in the ``Admin Dashboard''. 
The monitoring information is provided by a log processing system 
that extracts the relevant experiment task level execution logs from the 
gateway middleware and presents it in the gateway monitoring interfaces as depicted in Fig.~\ref{fig:fig5}.   

\subsection{Gateway Admin Dashboard}
The ``Admin Dashboard" is the workspace for the gateway administrator(s) within the gateway. 
All the admin features discussed above are available through the Admin Dashboard. 
Apart from what has been discussed, the admin dashboard provides a notification feature 
and also a way to managing gateway preferences when it comes to individual compute resource and 
storage resource connectivity and also helps managing credentials for secure compute resource 
communications. Gateway administrators can create notifications for gateway users and share 
them with the users with set begin and expiration times. The gateway admin can set and define 
preferences for the usage of each compute resource with specifications such as 
community or shared login name, job scratch location for the job data staging and execution, 
preferred job submission and file transfer protocols, and allocation project number for 
charging the run time. Similarly, the admin dashboard is used to generate an SSH credential 
token and key to be used  for authentication and authorization at the compute resource and storage resource communications. 
The admin dashboard also provides job scripts created by the Airavata middleware, the actual path of the job 
directory on the remote HPC system and detailed task level 
logs for a specific job to the administrator to check health of the job workflow or 
troubleshoot in case it is needed. 

\section{Community Building}
A proposal has been written to the MOLecular Software Sciences Institute(MOLSSI)~\cite{MOLSSI} 
to host a series of workshops designed to ;
\begin{enumerate}
\item Promote our ideas to a larger more diverse group of scientists than the ITAMP workshop participants 
both in A\&M physics as well as other related fields to help us solidify our ideas.
\item Initially conduct a three day workshop, most likely hosted by NIST sometime this fall.
\begin{itemize}
\item We envision inviting about 30 participants consisting of both A\&M scientists and quantum chemists.
\item We have requested support from the MOLSSI for the participants and have already 
been informed that we can expect \$15K to support our efforts.
\item The NSF Computational Physics program has also promised \$10K in support.
\end{itemize}
\item The workshop will consist of a number of general sessions discussing the codes  
and what they can and cannot do.  In addition, there will be hands on sessions 
on how to use the codes on the Science Gateway.
\item Develop a road-map for other workshops focusing more on a specific 
code or codes for specialists.
\end{enumerate}

\section{Acknowledgements}

BIS acknowledges the Mathematical Software group of the Applied and Computational Mathematics Division at NIST for supporting this work.  
This work used the Extreme Science and Engineering Discovery Environment (XSEDE) Stampede2, 
Comet, and Bridges resources, and an ECSS collaboration grant the through the 
allocation TG-PHY180023, which is supported by National Science Foundation grant number ACI-1548562. Additional support is 
provided by the Science Gateways Community Institute, NSF Award 1547611 \cite{Scigap2015}. 
The SciGaP.org platform and Apache Airavata are supported by NSF Award 1339774. KB and OZ acknowledge NSF support through grant 
Nos. PHY-1520970, PHY-1803844, and OAC-1834740. FM and MK acknowledge support by the ERC 
Proof-of-Concept Grant No. 780284-Imaging XCHEM within the Horizon 2020 Framework Program and by the Spanish MINECO Grant No. FIS2016-77889R. 
JT and JDG acknowledge support through the UK-AMOR
high end computing consortium under EPSRC grant EP/R029342/1.


\end{document}